\documentstyle[preprint,aps]{revtex}
\begin{document}
\preprint{UTPT-95-17}
\title{On Gravitational Collapse in the Nonsymmetric Gravitational Theory}
\author{J. W. Moffat and I. Yu. Sokolov}
\address{Department of Physics, University of Toronto,
Toronto, Ontario, Canada M5S 1A7}

\date{\today}

\maketitle

\begin{abstract}

\end{abstract}

The analytical structure of the difference between the static vacuum solution
in the
nonsymmetric gravitational theory (NGT) and the Schwarzschild solution of
Einstein's gravitational theory (EGT) is studied. It is proved that a smooth
matching
of the solutions does not exist in the range $0 < r \leq 2M$,  for any non-zero
values
of the parameters $M$ and $s$ of the NGT solution. This means that one cannot
consider the difference between the two solutions using perturbation theory in
this
range of $r$. Assuming that the exterior solution in gravitational collapse is
a small,
time dependent perturbation of the static solution for a non-zero, generic NGT
source ($s\not=0$) and mass density, it is shown that the matching of the
interior
and exterior solutions will not lead to black hole event horizons.


\narrowtext

In Einstein's gravitational theory (EGT), the collapse of a star leads
inevitably to
the formation of a black hole event horizon and a singularity at the center of
collapse
\cite{HawkingPenrose}. The event horizon is an infinite red shift surface in
any
coordinate frame of reference, which separates the spacetime manifold into two
causally disconnected pieces.

It has been conjectured that in the
gravitational collapse of a star in the nonsymmetric gravitational theory
(NGT),
a black hole event horizon will not form and the appearance of a singularity at
the center of collapse is not inevitable\cite{CornishMoffat1,CornishMoffat2}. A
detailed analysis of the gravitational collapse problem has been carried
out\cite{Moffat1}, using a new version of NGT which has a physically consistent
linear approximation without ghost poles, tachyons and possesses good
asymptotic
behavior\cite{Moffat2,Moffat3,LegareMoffat}. An analysis of the spherically
symmetric NGT system\cite{Moffat2,Clayton}, showed that a Birkhoff theorem
does not exist in NGT, i.e., the spherically symmetric vacuum solution is time
dependent.

Recently, Burko and Ori\cite{BurkoOri} have claimed that black holes can be
anticipated in gravitational collapse in NGT. They used the linear
approximation
for $g_{[\mu\nu]}$, expanded about the background Schwarzschild solution of
EGT.

In the following, we shall study the analytic properties of the static NGT
solution
and consider its consequences for the collapse of a star. This solution holds
in the
long-range approximation in which $\mu\approx 0$, where $\mu=1/r_0$ is a
``mass" parameter and the long-range approximation holds for large $r_0$. Since
we do not have a rigorous dynamical vacuum solution for NGT, we shall make the
reasonable physical assumption that
a quasi-static solution for the exterior of a star exists such that the limit
of this
solution to the static solution is smooth, i.e., the time dependent part of the
solution
is small and the dominant static piece determines the qualitative behavior of
the
solution\cite{Moffat1}. The NGT solution is a two-parameter static spherically
symmetric solution, in which the parameter $M$ is the mass and $s$ is a
dimensionless real parameter associated with the strength of the coupling of
$g_{[\mu\nu]}$ to matter. We can model $s$ by the expression\cite{Moffat1}:
\begin{equation}
\label{coupling}
s=\frac{g}{N^\beta},
\end{equation}
where $g$ is a coupling constant, $N$ denotes the particle number of a star and
$\beta$ is a dimensionless parameter. Thus, when $g$ is identically zero, the
NGT
vacuum solution reduces to the Schwarzschild solution of EGT.

In the case of a static spherically symmetric field, the canonical form of
$g_{\mu\nu}$
in NGT is given by
\begin{equation}
g_{\mu\nu}=\left(\matrix{-\alpha&0&0&w\cr
0&-\beta&f\hbox{sin}\theta&0\cr 0&-f\hbox{sin}\theta&
-\beta\hbox{sin}^2
\theta&0\cr-w&0&0&\gamma\cr}\right),
\end{equation}
where $\alpha,\beta,\gamma, f$ and $w$ are functions of $r$ and $t$. In the new
version of NGT, $w(r)$ does not satisfy the asymptotically flat boundary
conditions in the limit $r\rightarrow\infty$\cite{Clayton}. Therefore, in the
following, we shall set $w=0$ and only consider the unique two-parameter static
spherically symmetric solution for $f$, first obtained by Wyman\cite{Wyman}.

The NGT solution can be presented for $\beta=r^2$ as follows
\cite{CornishMoffat1,CornishMoffat2}:
\begin{mathletters}
\begin{eqnarray}
\label{gammaequation}
\gamma&=&e^\nu, \\
\alpha&=&\frac{M^2(\nu')^2 e^{-\nu}(1+s^2)}{(\cosh{(a\nu)}
-\cos{(b\nu)})^2},\\
\label{fequation}
f&=&\frac{2M^2e^{-\nu}[\sinh{(a\nu)}\sin{(b\nu)}
+s(1-\cosh{(a\nu)}\cos{(b \nu)})]}
{(\cosh{(a\nu)}-\cos{(b\nu)})^2},\\
a&=&\sqrt{{\sqrt{1+s^2}+1}\over 2},\\
b&=&\sqrt{{\sqrt{1+s^2}-1}\over 2}.
\end{eqnarray}
\end{mathletters}
Here, $\nu$ is given by the relation:
\begin{equation}
\label{nuequation}
e^\nu (\cosh{(a\nu)}-\cos{(b \nu)})^2 {r^2\over {2M^2}} =
\cosh{(a\nu)}\cos{(b \nu)}-1 +s\sinh{(a\nu)}\sin{(b\nu)}.
\end{equation}

For $2M/r \ll 1$ and $0 < sM^2/r^2 < 1$, the $\alpha, \gamma$ and $f$ take
the approximate forms ($\mu^{-1} \gg 2M$):
\begin{equation}
\gamma\approx\alpha^{-1}\approx 1-\frac{2M}{r},\, f\approx \frac{sM^2}{3}.
\end{equation}
This result guarantees that all the experimental tests of EGT, based on the
exterior
point source Schwarzschild solution, are valid for large $r$ and a suitably
chosen
value of the parameter $s$.

In order to study the analytical structure of the static Wyman solution, we
shall
consider the expansion of the solution in a power series in $s$ about the
Schwarzschild solution. Namely, we need to represent the
nonsymmetric $g_{\mu\nu}$ solution in the following way:
\begin{equation}
\label{difference}
g_{\mu \nu}=g_{\mu \nu}^S + \Delta_{\mu \nu},
\end{equation}
where $g_{\mu \nu}^S$ is the Schwarzschild metric tensor and $\Delta_{\mu \nu}$
is the sought difference. The points where such an expansion does not exist
will be
specific points at which the NGT solution is a non-analytic function of
the parameter $s$.

Let us consider first the difference $\Delta_{\mu \nu}$ inside the horizon
($r\leq
2M$). As was shown in\cite{CornishMoffat1,CornishMoffat2}, the diagonal
components of $g_{\mu \nu}$ do not change their signs when crossing the point
$r=2M$. Such a behavior does not depend on
the parameter $s$. At the same time, the metric tensor
$g^S_{\mu \nu}$ changes sign when crossing
the horizon. It means that it is impossible to realize the
sought expansion of the NGT solution near the Schwarzschild solution for $r\leq
2M$.
However, such an expansion exists when we consider
the analytical continuation of diagonal elements of $g_{\mu \nu}$ to negative
values.
This continuation exists on the field of complex numbers. In order to  perform
the
continuation, it is enough to consider $g_{\mu\nu}$ as a complex-valued
function,
keeping the radius $r$, mass $M$ and the constant $s$ real.

We shall demonstrate the technique of analytical continuation
for the example of the $g_{44}$-component of $g_{\mu \nu}$.
It is seen from Eq.(\ref{gammaequation}) that to have a negative $\gamma$, one
should consider a pure imaginary $\nu$. For example, if $s=0$ (Schwarzschild
solution), it leads to a real negative $\gamma$ for $r<2M$. Provided $s\not=0$
(NGT solution), the continuation for  $\gamma$ becomes a complex valued
function. To find this analytical continuation, one needs to use the known
equality
\[
\ln \gamma = \ln |\gamma| +{\rm i}(2\pi k + \varphi ),
\]
where $\varphi$ is the phase of $\gamma$ and $k$ is an integer number.
Then, one sees that the continuation will be a multivalued
function. Hereafter, for the sake of simplicity, let us put $k=0$.

Now one may find the continuation by solving Eq.(\ref{nuequation}) with respect
to
complex $\nu$ (the connection between $\nu$ and $\gamma$ is given by
Eq.(\ref{gammaequation})). Regrettably, there is no explicit solution of this
equation.
Let us recall, however, that the region of interest is near $s=0$. So, we shall
consider
the analytical continuation of $\gamma$ for small $s$. Expanding  $\gamma$ in
a power series in $s$:
\begin{equation}
\label{ser}
\gamma = c_0 + c_1 s + c_2 s^2 + \dots
\end{equation}
and substituting (\ref{difference}) into (\ref{nuequation}),
we can find the equations for the coefficients $c_0,\ c_1 \ \dots$. As a
result, we get
\begin{equation}
\label{delta_4}
\Delta_{44}=\gamma -1+\frac{2m}{r}=-(d_1+{\rm i}\ d_2)s^2+O[s^4],
\end{equation}
where
\begin{eqnarray}
d_1&=&\frac{\pi^2}{8}\;\biggl(\frac{M}{r}\biggr)\,\biggl[
1-3\biggl(\frac{r}{M}\biggr)
+\frac{3}{2}\biggl(\frac{r}{M}\biggr)^2+
\biggl(1+3\biggl(\frac{r}{M}\biggr)-6\biggl(\frac{r}{M}\biggr)^2\biggr)
\ln\biggl(\frac{2M}{r}-1\biggr)\biggr],\\
d_2&=&\frac{\pi}{8}\;\biggl(\frac{M}{r}\biggr)\,\biggl[3
-3\biggl(\frac{r}{M}\biggr)
+\biggl(-2+6\biggl(\frac{r}{M}\biggr)-3\biggl(
\frac{r}{M}\biggr)^2\biggr) \ln\biggl(\frac{2M}{r}-1\biggr)\biggr].
\end{eqnarray}

It should be stressed that we have made no assumptions about
the value of $r$ apart from $r \leq 2M$. Therefore, the expansion
(\ref{delta_4})
must be valid for any $r \leq 2M$. One can then see that the continuation
does not exist at the two points $r=0$ and $r=2M$, i.e., at the points of the
origin and the horizon (logarithmic features).
This behavior has been confirmed by an exact (numerical) calculation.

The same conclusion can be obtained for
$\alpha$ and $f$. Using the aforementioned technique, we find for the
difference between $\alpha^{-1}$ for NGT and the Schwarzschild solution:
\begin{eqnarray}
\alpha^{-1} - (1 - \frac{2M}{r})=\frac{Ms^2}{16r}\biggl[
-44+28\frac{M}{r} +\pi^2 \biggl(6- 24\frac{M}{r}
+25\biggl(\frac{M}{r}\biggr)^2-7\biggl(\frac{M}{r}\biggr)^3\biggr)
\nonumber \\
+\biggl(22 -24\biggl(\frac{M}{r}\biggr)-22\biggl(\frac{M}{r}\biggr)^2
+14\biggl(\frac{M}{r}\biggr)^3\biggr)
\ln{\biggl(\frac{2M}{r} -1\biggr)}
\nonumber \\
+{\rm i}\pi\biggl\{34 -72\biggl(\frac{M}{r}\biggr)
+28\biggl(\frac{M}{r}\biggr)^2
+
(-12 +48\biggl(\frac{M}{r}\biggr)-50\biggl(\frac{M}{r}\biggr)^2\nonumber \\
+14\biggl(\frac{M}{r}\biggr)^3\biggr)\ln\biggl(\frac{2m}r -1\biggr)
\biggr\}\biggr]+O[s^4],
\end{eqnarray}
and
\begin{equation}
f\equiv{\Delta_{23}\over \sin{\theta}}={-\Delta_{32}\over \sin{\theta}}=
-\frac{1}{2}r^2 s\biggl[2-\biggl(\frac{r}{M}
-1\biggr)\ln\biggl(\frac{2m}{r}-1\biggr)
+{\rm i}\pi\biggl(\frac{r}{M}-1\biggr)\biggr] + O[{s}^5].
\end{equation}
It should be noted that, in contrast to the other components of the analytical
continuation of $g_{\mu\nu}$, the function $f$ is regular at $r=0$.

Now we shall consider the analytical structure of the difference
beween the NGT and Schwarzschild solutions with $r\geq 2M$.
There is no point in perfoming the continuation
in the way that was done before, because the difference exists
in real numbers. We can find it by substituting (\ref{ser})
into Eqs.(\ref{gammaequation})-(\ref{fequation}). The result of the calculation
is given by
\begin{eqnarray}
\Delta_{44}&=&\frac{rs^2}{16 M [-4 +3 \left(\frac{r}{M}\right)^2]
\ln\biggl(1-\frac{2M}{r}\biggr)} + O[s^4],\\
\alpha^{-1}-1+{2M\over r}&=& \frac{Ms^2}{8r}
\biggl[22-14\frac{r}{M}\biggl(-11+12\biggl({r\over M}\biggr)\nonumber\\
&&+6\biggl({r\over M}\biggr)^2
-7\biggl({r\over M}\biggr)^3 \biggr)\ln{\biggl(1-{2M \over r}\biggr)}\biggr]
+O[s^4], \\
f&\equiv&{\Delta_{23}\over \sin{\theta}}\nonumber \\
&=&{-\Delta_{32}\over \sin{\theta}}=
\frac{1}{2} r^2 s\biggl[\biggl(1-\frac{r}{M}\biggr)\ln{\biggl(1-{2M \over r}
\biggr)}-2\biggr] +O[s^3].
\end{eqnarray}

It should be noted that the asymptotics of these equations for
$r/M \gg 2$ are in agreement with those presented in
Refs.\cite{CornishMoffat1,CornishMoffat2}.
This result can also be corroborated by numerical calculations.

We reach the following conclusion: the difference between the NGT solution
and the Schwarzschild solution is a regular function,
complex-valued in the open range $0< r/M <2$ and real-valued
in the range $r/M > 2$, it has a non-analytic logarithmic behaviour near
$r/M=0,2$,
i.e., at the origin and at the Schwarzschild horizon. In particular, this means
that it is impossible to match smoothly the Schwarzschild and NGT solutions in
the
neighborhood of $r/M=0,2$ for any value of the parameter $s\not=0$. A small
first
order static $g_{[\mu\nu]}$ on a Schwarzschild background {\it is not a global
solution
of the NGT static vacuum field equations}. Moreover, if the time dependent part
of
$g_{[\mu\nu]}$ is small and is a smooth function on a Schwarzschild background,
then this is not expected to be a global solution either for $s\not=0$.

We must now consider the matching of the interior and exterior solutions during
the
collapse of a star. The interior and exterior solutions match at the surface of
the star,
$r=r_0$, if we have\cite{Moffat1}
\begin{mathletters}
\begin{eqnarray}
\gamma(r_0,t)&=&\gamma_{\rm ext}(r_0,t),\\
\alpha(r_0,t)&=&\alpha_{\rm ext}(r_0,t),\\
\beta(r_0,t)&=&\beta_{\rm ext}(r_0,t),\\
f(r_0,t)&=&f_{\rm ext}(r_0,t),
\end{eqnarray}
\end{mathletters}
where $\gamma_{\rm ext}, \alpha_{\rm ext}, \beta_{\rm ext}$ and $f_{\rm ext}$
denote the non-vanishing components of $g_{\mu\nu}(r,t)$ for the exterior
time dependent vacuum solution of the NGT field equations.

We shall expand the exterior time dependent solution as
\begin{mathletters}
\begin{eqnarray}
\gamma_{\rm ext}(r,t)&=&\gamma_{\rm ext}(r)+\delta\gamma_{\rm ext}(r,t),\\
\alpha_{\rm ext}(r,t)&=&\alpha(r)_{\rm ext}+\delta\alpha_{\rm ext}(r,t),\\
\beta_{\rm ext}(r,t)&=&\beta_{\rm ext}(r)+\delta\beta_{\rm ext}(r,t),\\
f_{\rm ext}(r,t)&=&f_{\rm ext}(r)+\delta f_{\rm ext}(r,t).
\end{eqnarray}
\end{mathletters}
Because we do not have an exact dynamical solution of the vacuum NGT field
equations, let us assume that $\delta\gamma_{\rm ext}, \delta\alpha_{\rm ext},
\delta\beta_{\rm ext}$ and $\delta f_{\rm ext}$ are small quantities that can
be
neglected,
without encountering any discontinuities when going to limit of the static part
of
$g_{\mu\nu}$. We shall refer to this
as the quasi-static approximation of the vacuum field equations.

We have assumed that there exists a non-vanishing generic coupling of
$g_{[\mu\nu]}$
to the matter composing the star and that there is no NGT neutral body in
nature,
i.e., the coupling (\ref{coupling}) is always non-zero in the presence of a
matter
source. This guarantees the existence of a static exterior $g_{\mu\nu}$ which
is given for a spherically symmetric star by the Wyman solution. Burko and
Ori\cite{BurkoOri} assumed that $f(r,t)$ was small at the beginning of the
collapse
of a star and {\it continued to be small of first order throughout the
collapse}
(with the possible exception of $r=0$). Due to the assumption of the smallness
of $f(r,t)$, they were implicitly assuming that the static Schwarzschild
background
solution dominated the collapse for $r \leq 2M$, i.e., they assumed that the
exterior
metric was described by a quasi-static approximation. We have proved
that this cannot be a global solution of the NGT field equations; in
particular, it fails
to be a solution for $0 < r \leq 2M$.

{}From the static Wyman solution, we know that $f(r)$ has to
be greater than 1 in Cartesian coordinates for
$r\sim 2M$\cite{CornishMoffat1,CornishMoffat2}. Therefore, according to our
approximation scheme, the quasi-static solution for $f$ is also expected to
give $f > 1$ in Cartesian coordinates. It follows that the claim by
Burko and Ori that black holes can form in NGT for
small enough $f$ fails to be true, for it is based on the validity of
the linear approximation equation\cite{BurkoOri}:
\begin{equation}
\label{flinear}
\frac{1}{2}\biggl(\frac{\ddot{f}}{\gamma}-\frac{f''}{\alpha}\biggr)
+\frac{f'}{\alpha
r}+\frac{1}{2}\frac{f'\alpha'}{\alpha^2}-2\frac{f\alpha'}{\alpha^2r}
=0,
\end{equation}
which reduces in the limit $\dot{f}\rightarrow 0$ to the static equation for
$f$\cite{Clayton,Cornish}. The same arguments hold for the form of
Eq.(\ref{flinear}) in Kruskal-Szekeres coordinates\cite{BurkoOri}.

An analogous situation holds in EGT, if we expand the metric tensor about
Minkowski flat space. To first order we obtain the metric:
\[
ds^2=(1-\frac{2M}{r})dt^2-(1+\frac{2M}{r})dr^2-r^2(d\theta^2
+\sin^2\theta d\phi^2).
\]
As the star collapses and $r\rightarrow 2M$, the linear expansion breaks
down and we are required to solve the non-linear EGT field equations.

It is possible that there exists one exotic situation, namely, that there is no
static part
to $f$, i.e., $f$ is a purely source-free
wave solution of the NGT field equations. Since we do not have a rigorous
wave-type solution of the NGT field equations, we cannot at present know
whether
such a
solution correctly describes the collapse problem near the Schwarzschild
radius,
$r\sim 2M$, when it is restricted to the linear approximation. A numerical
solution
of the NGT field equations may shed some light
on this question. However, we expect that for a non-vanishing generic coupling
of $f$ to matter, a static part of $f$ should exist for realistic collapse, and
therefore
our arguments about the non-formation of black holes would follow. We
anticipate
that black holes do not form, in NGT, for a general dynamical solution of the
field equations.

\acknowledgements

This work was supported by the Natural Sciences and Engineering Research
Council of Canada. We thank M. A. Clayton, N. J. Cornish, L. Demopoulos
and M. Reisenberger for helpful discussions.

\end{document}